\documentclass[twocolumn,showpacs,prb,superscriptaddress]{revtex4}

\usepackage{graphicx}

\newcommand{\av}{_{\mathrm{av}}}
\newcommand{\nsw}{N_{\mathrm{sweep}}}
\newcommand{\nsa}{N_{\mathrm{samp}}}
\newcommand{\figurewidth}{\columnwidth}

\begin{document}

\title{Correlation length of the two-dimensional Ising spin glass with 
Gaussian interactions}

\author{Helmut G.~Katzgraber}
\affiliation
{Theoretische Physik, ETH H\"onggerberg,
CH-8093 Z\"urich, Switzerland}

\author{L.~W.~Lee}
\email{leelikwe@physics.ucsc.edu}
\affiliation{Department of Physics,
University of California,
Santa Cruz, California 95064, USA}

\author{A.~P.~Young}
\email{peter@bartok.ucsc.edu}
\homepage{http://bartok.ucsc.edu/peter}
\affiliation{Department of Physics,
University of California,
Santa Cruz, California 95064, USA}

\date{\today}

\begin{abstract}
We study the correlation length of the two-dimensional Ising spin glass with
a Gaussian distribution of interactions, using
an efficient Monte Carlo algorithm proposed by Houdayer, that allows larger
sizes and lower temperatures to be studied than was possible
before. We find that the
``effective'' value of the bulk correlation length exponent $\nu$ increases as
the temperature is lowered, and, at low temperatures, apparently approaches 
$-1/\theta$, where $\theta \simeq -0.29$ is the stiffness exponent obtained 
at zero temperature. This means scaling is satisfied and earlier results 
at higher temperatures that find a smaller value for $\nu$ are affected 
by corrections to scaling.
\end{abstract}

\pacs{75.50.Lk, 75.40.Mg, 05.50.+q}
\maketitle

\section{Introduction}
\label{sec:introduction}

There are two main theories to describe the spin-glass state: the droplet
theory\cite{fisher:86,fisher:88,bray:86,mcmillan:84a} and the replica symmetry
breaking theory\cite{parisi:79,parisi:80,parisi:83,mezard:87} (RSB) of Parisi.
According to the droplet picture the lowest energy excitation, or ``droplet,''
of linear size $l$ containing a given site has a characteristic energy of
order $l^\theta$ where $\theta > 0$ is a ``stiffness'' exponent.  Droplets are
expected to be compact but with a surface that has a nontrivial fractal
dimension $d_s$, less than the space dimension $d$. It is further assumed that
the same exponent $\theta$ describes both droplet and ``domain-wall''
excitations.  In the alternative RSB scenario, the energy of droplets
containing a finite fraction of the system does not increase with increasing
system size.  Furthermore, in RSB the surface of the large-scale, low-energy
excitations are expected to fill space and so have the fractal dimension $d_s
= d$. 

There have been many numerical studies in
three\cite{bray:84,mcmillan:84,hartmann:99,palassini:99,marinari:98a}
and four\cite{hartmann:99a,hukushima:99} dimensions
that attempt to determine which of
these scenarios, or possibly something else\cite{krzakala:00,palassini:00}, is
correct. These calculations are quite limited in the range of sizes that can
be studied, although recently\cite{katzgraber:03,katzgraber:03f}
a larger range has been studied for a one-dimensional model with power law
interactions. Another case where a large range of sizes can be studied is
the two-dimensional spin glass with short-range interactions, for which there
is no spin-glass order at finite temperature, corresponding to $\theta <
0$. It is desirable to understand fully the two-dimensional spin glass,
including the nature of corrections to scaling, since this may help in the
interpretation of numerical data in higher dimensions.

However, even in $d=2$, the situation is not completely clearcut.
Zero-temperature calculations of the energy of a domain
wall\cite{mcmillan:84b,bray:84,rieger:96,palassini:99a,hartmann:01a,hartmann:02c}
consistently give $\theta \simeq-0.29$, with Ref.~\onlinecite{hartmann:02c},
for example, quoting $\theta = -0.287 \pm 0.004$.
However, some calculations of droplet energies
find\cite{kawashima:00a,kawashima:00} $\theta \simeq -0.47$, while 
others\cite{hartmann:02} find results consistent with the domain-wall value.
These discrepancies presumably arise because some of the results are
affected by corrections to scaling, and, as discussed by Hartmann and
Moore,\cite{hartmann:03} the domain-wall value, $\theta \simeq -0.29$, seems to
be the correct asymptotic result.

Although the discrepancy between the estimates for $\theta$ from
the zero-temperature calculations seems now to be resolved,\cite{hartmann:03} 
there is still an
apparent conflict between $\theta$ and the value of the correlation length
exponent $\nu = 2.0 \pm 0.2$ obtained from finite temperature Monte Carlo
simulations.\cite{liang:92} Since the spin-glass transition occurs at $T=0$ in
two dimensions, the correlation length $\xi$ diverges as $ T \to 0$ as
$\xi \sim T^{-\nu}$. According to scaling,\cite{bray:84,mcmillan:84}
$\nu$ is related to $\theta$ by 
\begin{equation}
\nu = -{1 \over \theta} \, ,
\label{eq:nu}
\end{equation}
which gives $\nu \simeq 3.5$,
significantly different from the result
$\nu = 2.0 \pm 0.2$
reported in Ref.~\onlinecite{liang:92}. We investigate this discrepancy here
by performing Monte Carlo simulations on the Ising spin glass with Gaussian
interactions in two dimensions at larger sizes and lower temperatures than in
Ref.~\onlinecite{liang:92}. 

\section{Model and Observables}
\label{sec:model}

The Hamiltonian is
\begin{equation}
{\cal H} = - \sum_{\langle i,j\rangle} J_{ij} S_i S_j ,
\label{eq:ham}
\end{equation}
where the sum is over nearest neighbor pairs of sites on a square lattice,
the $S_i$ are Ising spins taking values $\pm 1$, and the $J_{ij}$ are Gaussian
variables with zero mean and standard deviation unity. The square lattice contains $N
= L \times L$ sites with periodic boundary conditions. We use the Monte Carlo
algorithm of Houdayer\cite{houdayer:01} which combines single spin flip
dynamics, parallel tempering,\cite{hukushima:96} and a type of cluster move,
which is significantly more efficient than parallel tempering in two
dimensions for large system sizes ($L > 24$). 
Tests for equilibration are done as in
Ref.~\onlinecite{katzgraber:01}; the parameters used in the simulations are 
shown in Table \ref{simparams}.

\begin{table}
\caption{
Parameters of the simulations. $\nsa$ is the number of samples,
$\nsw$ is the total number of Monte Carlo sweeps for each of the $2 N_T$
replicas for a single sample, $T_{\rm min}$ is the lowest temperature 
simulated, and $N_T$ is the number of temperatures used
in the parallel tempering method. Note that for $L \le 16$ standard parallel 
tempering Monte Carlo was used, whereas for $L \ge 32$ the cluster
method by Houdayer was applied.
\label{simparams}
}
\begin{tabular*}{\columnwidth}{@{\extracolsep{\fill}} c r r r l }
\hline
\hline
$L$  &  $\nsa$  & $\nsw$ & $T_{\rm min}$ & $N_{T}$  \\
\hline
  8 & $10000 $ & $2.0 \times 10^5$ & 0.05 & 20 \\
 16 & $10000 $ & $1.0 \times 10^6$ & 0.05 & 20 \\
 32 & $10000$ & $1.0 \times 10^5$ & 0.05 & 20 \\
 64 & $1000$ & $1.0 \times 10^6$ & 0.05 & 40 \\
128 & $250$ & $1.0 \times 10^6$ & 0.20 & 63 \\
\hline
\hline
\end{tabular*}
\end{table}

The main focus of our study is the correlation length \textit{of the finite
system}\cite{kim:94,palassini:99b,ballesteros:00,lee:03} $\xi_L$, defined by
\begin{equation}
\xi_L = {1 \over 2 \sin (|{\bf k}_\mathrm{min}|/2)}
\left[{\chi_{\mathrm{SG}}(0) \over 
\chi_{\mathrm{SG}}({\bf k}_\mathrm{min})} - 1
\right]^{1/2},
\label{eq:xiL}
\end{equation}
where ${\bf k}_\mathrm{min} = (2\pi/L, 0, 0)$ is the smallest nonzero
wave vector, and
\begin{equation}
\chi_{\mathrm{SG}}({\bf k}) = {1 \over N} \sum_{i,j} [\langle
S_i S_j
\rangle^2 ]\av e^{i {\bf k}\cdot({\bf R}_i - {\bf R}_j) } 
\label{eq:chik}
\end{equation}
is the wave vector-dependent spin-glass susceptibility. In Eq.~(\ref{eq:chik})
the angular brackets $\langle \cdots \rangle$ denote a thermal average while
the rectangular brackets $[ \cdots ]\av$ denote an average over the disorder.

Since the ratio $\xi_L/L$ is dimensionless, it satisfies the finite size
scaling form\cite{palassini:99b,ballesteros:00,lee:03}
\begin{equation}
{\xi_L / L } = \tilde{X} [ L^{1/\nu} T ] \, ,
\label{eq:fss}
\end{equation}
assuming a zero-temperature transition, where $\tilde{X}$ is a scaling 
function and $1/\nu = -\theta$.  For
$ L^{1/\nu} T \gg 1$, $\xi_L$ is equal to the bulk (i.e., infinite system size)
correlation length $\xi_\infty$, and so
\begin{equation}
\xi_L = \xi_\infty \sim T^{-\nu} \qquad (L \gg \xi_\infty) \, ,
\label{eq:xi}
\end{equation}
implying that $\tilde{X}(x) \sim x^{-\nu}$ for $x \gg 1$. In the opposite
limit, $x \ll 1$, we expect $\tilde{X}(x) \sim x^{-\lambda}$ where we will 
estimate $\lambda$ below.

\begin{figure}
\includegraphics[width=\figurewidth]{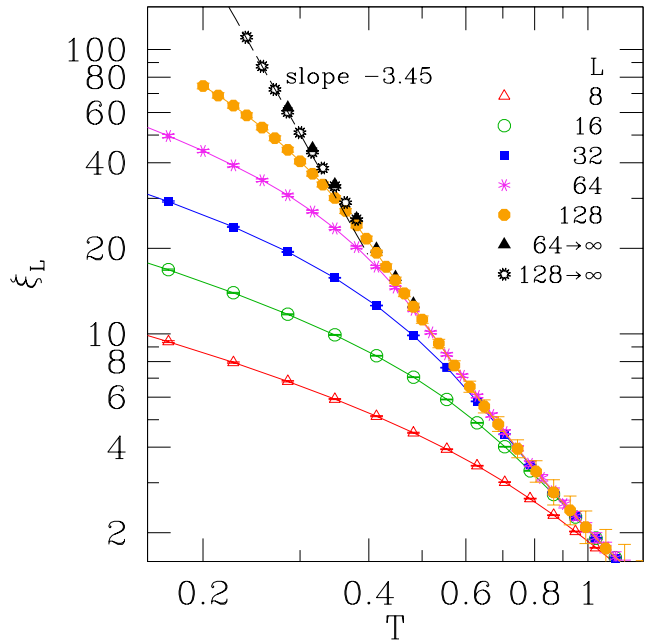}
\caption{
A log-log plot of the
finite size correlation length $\xi_L$ for different system sizes and
temperatures. The data labeled $64\to\infty$ and $128\to\infty$ come from
an extrapolation to the thermodynamic limit. The slope
of the extrapolated data gives $-\nu \simeq -3.45$, which is consistent with
$\theta \equiv -1/\nu
\simeq -0.287$ found in zero-temperature domain-wall
calculations (Ref.~\onlinecite{hartmann:02c}).
}
\label{xi}
\end{figure}

\section{Results}
\label{results}

Figure \ref{xi} shows data for $\xi_L$. We see that the data are independent
of system size at high $T$, showing that the bulk behavior has been obtained, 
and the data ``peels off'' from this ``bulk curve'' at temperatures that
become 
successively lower for larger sizes. The bulk results are curved on this
log-log plot showing that the asymptotic power law behavior in
Eq.~(\ref{eq:xi}) has not yet been reached. Rather, the slope of the curve is
an ``effective exponent,'' $\nu_{\rm eff}$, which varies with $T$.

In order to determine the asymptotic value of $\nu$ we obtain values for the
bulk correlation length at lower $T$ following the method used by
Kim.\cite{kim:94} The finite-size scaling expression,
Eq.~(\ref{eq:fss}), can be written as
\begin{equation}
{\xi_L \over L} = f\left({\xi_\infty \over L}\right) \, ,
\end{equation}
which can be inverted to give
\begin{equation}
{\xi_\infty \over L} = g\left({\xi_L \over L}\right) \, ,
\label{gx}
\end{equation}
where $g(x) = f^{-1}(x)$. Clearly
$g(x) = x$ for $x \to 0$.
We determine $g(x)$ by fitting to data in the range
$0.45 < T < 1.05$ where we have data for the correlation length in
\textit{both} the bulk and finite-size regimes.  We consider sizes $16 \le L
\le 128$ for this determination, from which we obtain
$g(x)$ in the range $0 < x <
0.45$. Using $g(x)$ \textit{in this range} we then 
determine $\xi_\infty$ from Eq.~(\ref{gx}) using
data for $L=64$ in the range $ 0.285  \le T \le 0.482$ and for
$L=128$ in the range $ 0.24 \le T \le 0.38$. Note, that we do not perform any
direct extrapolation
in this analysis; the function $g(x)$ is determined by fitting
and is then used to get $\xi_\infty$ at somewhat lower temperature
using only the range of $x$ where it was fitted.

The resulting values of $\xi_\infty$ from the $L=64$ and 128 data are
consistent with each other where they overlap,
and are shown in Fig.~\ref{xi}.
The extrapolated
data fit a slope of $-\nu = -3.45$
which corresponds to $\theta \equiv -1/\nu = -0.29$ in
good agreement with domain-wall results.\cite{hartmann:02c} 
At higher temperature, the slope of the bulk data in Fig.~\ref{xi} is
smaller in magnitude, so we believe that Liang's result, $\nu = 2.0 \pm 0.2$,
obtained in the region around $T = 1.0$,
is only an effective exponent.
Neither our
results nor the domain-wall results appear to be consistent with an exponential
divergence of the correlation length as $T \to 0$.

Assuming that the asymptotic value of $\nu$ is indeed $\approx 3.45$
we estimate from
Fig.~\ref{xi} that one needs to be at or below a temperature of around $0.35$,
where $\xi_\infty \simeq 40$, to see the bulk asymptotic behavior.

\begin{figure}
\includegraphics[width=\figurewidth]{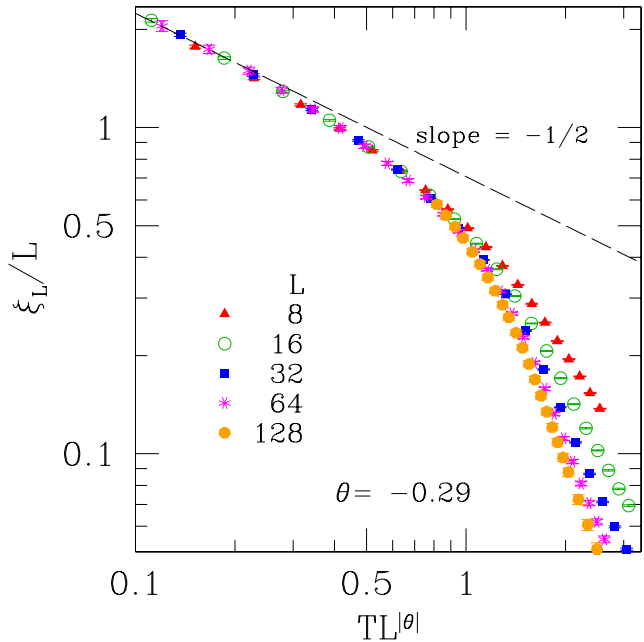}
\caption{
Scaling plot of the data for the correlation length $\xi_L$ according to 
Eq.~(\ref{eq:fss}), with $\theta = -1/\nu = -0.29$. The dashed line, which 
fits the data at low $T$, has slope $-1/2$.
}
\label{xi_scale_0.29}
\end{figure}

Further insight is obtained by scaling the data according to
Eq.~(\ref{eq:fss}). Figure \ref{xi_scale_0.29} shows a scaling
plot with $\theta =
-0.29$, the value expected from zero-temperature domain-wall calculations and
our extrapolated data in Fig.~\ref{xi}. We
see the data in Fig.~\ref{xi_scale_0.29}
scale well at low $T$ but not well at high $T$. However,
the latter point is not surprising in view of Fig.~\ref{xi} where we see that
asymptotic bulk power-law behavior has not yet been reached. Note, though,
that the data for the two largest sizes
in Fig.~\ref{xi_scale_0.29}, $L=64$ and 128,
do almost collapse, suggesting that
$\theta = -1/\nu = -0.29$ \textit{will}
work in the bulk region, $L \gg \xi$, for large
enough sizes and low enough temperatures, as we also inferred from
Fig.~\ref{xi}.

The dashed line in Fig.~\ref{xi_scale_0.29} has slope $-1/2$, implying from
Eq.~(\ref{eq:fss}) that $\tilde{X}(x) \sim x^{-\lambda}$ for $x \ll 1$ 
with $\lambda \simeq 1/2$. Hence, we have
\begin{equation}
\xi_L \sim T^{-1/2} L^{1-1/(2\nu)} \qquad (L \ll \xi) \, .
\label{eq:xilowT}
\end{equation}
The $T^{-1/2}$ dependence can be understood from Eq.~(\ref{eq:xiL}) since
$\chi_{\mathrm{SG}}(0) = L^2$ at $T = 0$ (because the ground state is
unique), while the fluctuations at
nonzero ${\bf k}$ are frozen out at $T = 0$. It is plausible
$\chi_{\mathrm{SG}}({\bf k}_\mathrm{min}) \propto T$
at low $T$ from equipartition, and
this leads to a $T^{-1/2}$ dependence for $\xi_L$.

\begin{figure}
\includegraphics[width=\figurewidth]{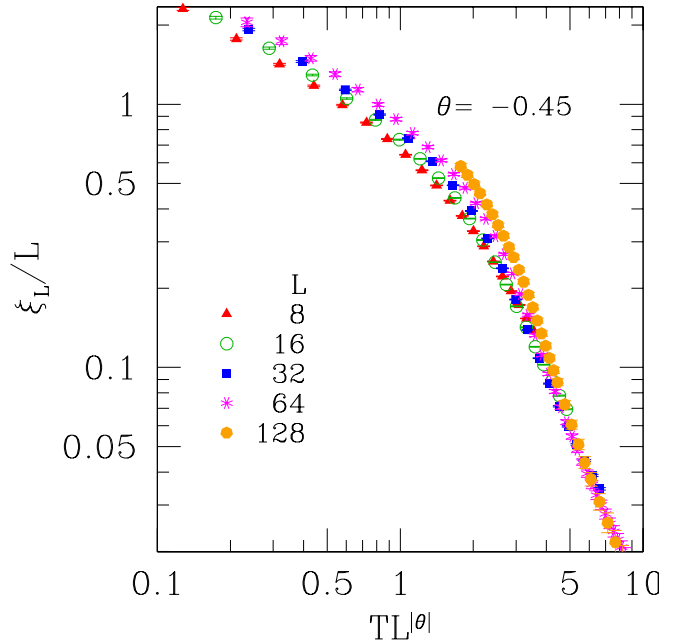}
\caption{
Scaling plot of the data for the correlation length $\xi_L$ according to 
Eq.~(\ref{eq:fss}), with $\theta = -1/\nu = -0.45$. 
}
\label{xi_scale_0.45}
\end{figure}

Figure \ref{xi_scale_0.45} shows a scaling plot for $\theta = -0.45$, which
gives the best data collapse in the high-$T$ region. This value is compatible
with $\nu = 2.0 \pm 0.2$ found in Ref.~\onlinecite{liang:92}. Note, that
the data do
not collapse at all in the low-$T$ region and the collapse becomes worse for
larger sizes.
Figure \ref{xi_scale_0.45} shows, again, that an \textit{effective} value of 
$\nu \approx 2$ will fit the data over a range of intermediate temperatures, 
while in the low-$T$ asymptotic region one has $\theta = -1/\nu
\simeq 0.29$.
We have found that the 
spin glass susceptibility shows similar behavior. Independent 
recent results for the spin
glass susceptibility and other quantities~\cite{houdayer:04}
also find evidence that $\theta \simeq -0.29$ for large sizes.

\section{Conclusions}
\label{conclusions}

To conclude, we have shown that the result $\nu = 2.0 \pm 0.2$ in
Ref.~\onlinecite{liang:92} is only an effective exponent, and the true value
for $\nu$ is larger. The data are consistent with scaling holding
asymptotically for $\nu = -1/\theta \simeq 3.45$. We have strengthened our
argument for this conclusion
by the extrapolation to $L=\infty$ shown in Fig.~\ref{xi}.
Of course it would also be desirable to extend the data to larger sizes, which
may be possible in the near future by fine tuning the
algorithm.

\begin{acknowledgments}
The authors would like to thank J.~Houdayer and A.~Hartmann 
for helpful discussions and
for sending us a copy of their work (Ref.~\onlinecite{houdayer:04}) prior 
to submission.
We would also like to thank M.~Troyer for pointing out the extrapolation
method introduced in Ref.~\onlinecite{kim:94}.
L.W.L.~and A.P.Y.~acknowledge support
from the National Science Foundation under NSF Grant No.~DMR 0337049.
The simulations were performed on the Asgard cluster
at ETH Z\"urich.
\end{acknowledgments}

\bibliography{refs}

\end{document}